\documentclass[aps,prc,twocolumn,superscriptaddress,showpacs]{revtex4-1}
\usepackage{graphicx}
\usepackage{dcolumn}
\usepackage{bm}
\usepackage{slantsc}%
\usepackage{subfigure}
\usepackage{amssymb}
\usepackage{MnSymbol}
\newcommand{\beq}{\begin{equation}}
\newcommand{\eeq}{\end{equation}}
\newcommand{\beqar}{\begin{eqnarray}}
\newcommand{\eeqar}{\end{eqnarray}}
\newcommand{\ds}{\displaystyle}
\newcommand{\bec}{\begin{center}}
\newcommand{\enc}{\end{center}}
\usepackage{color}

\begin{document}

\title{ Influence of jets and decays of resonances on the
triangular flow in ultrarelativistic heavy-ion collisions }

\author{J.~Crkovsk\'{a}}
\affiliation{
Faculty of Nuclear Sciences and Physical Engineering, Czech Technical 
University in Prague, CR-11519 Prague, Czech Republic
\vspace*{1ex}}
\affiliation{
Institut de Physique Nucl\'{e}aire, CNRS-IN2P3, Univ. Paris-Sud, 
Universit\'{e} Paris-Saclay, F-91406 Orsay Cedex, France
\vspace*{1ex}}
\author{J.~Biel\v{c}\'{i}k}
\affiliation{
Faculty of Nuclear Sciences and Physical Engineering, Czech Technical 
University in Prague, CR-11519 Prague, Czech Republic
\vspace*{1ex}}
\author{L.~Bravina}
\affiliation{
Department of Physics, University of Oslo, PB 1048 Blindern,
N-0316 Oslo, Norway
\vspace*{1ex}}
\affiliation{
Frankfurt Institute for Advanced Studies, Ruth-Moufang-Stra{\ss}e 1,
D-60438 Frankfurt a.M., Germany
\vspace*{1ex}}
\affiliation{
National Research Nuclear University "MEPhI" (Moscow Engineering 
Physics Institute), RU-115409 Moscow, Russia
\vspace*{1ex}}
\author{B.H.~Brusheim~Johansson}
\affiliation{
Department of Physics, University of Oslo, PB 1048 Blindern,
N-0316 Oslo, Norway
\vspace*{1ex}}
\affiliation{
Frankfurt Institute for Advanced Studies, Ruth-Moufang-Stra{\ss}e 1,
D-60438 Frankfurt a.M., Germany
\vspace*{1ex}}
\author{E.~Zabrodin}
\affiliation{
Skobeltsyn Institute of Nuclear Physics,
Moscow State University, RU-119991 Moscow, Russia
\vspace*{1ex}}
\affiliation{
Department of Physics, University of Oslo, PB 1048 Blindern,
N-0316 Oslo, Norway
\vspace*{1ex}}
\affiliation{
Frankfurt Institute for Advanced Studies, Ruth-Moufang-Stra{\ss}e 1,
D-60438 Frankfurt a.M., Germany
\vspace*{1ex}}
\affiliation{
National Research Nuclear University "MEPhI" (Moscow Engineering 
Physics Institute), RU-115409 Moscow, Russia
\vspace*{1ex}}
\author{G.~Eyyubova}
\affiliation{
Skobeltsyn Institute of Nuclear Physics,
Moscow State University, RU-119991 Moscow, Russia
\vspace*{1ex}}
\author{V.L.~Korotkikh}
\affiliation{
Skobeltsyn Institute of Nuclear Physics,
Moscow State University, RU-119991 Moscow, Russia
\vspace*{1ex}}
\author{I.P.~Lokhtin}
\affiliation{
Skobeltsyn Institute of Nuclear Physics,
Moscow State University, RU-119991 Moscow, Russia
\vspace*{1ex}}
\author{L.V.~Malinina}
\affiliation{
Skobeltsyn Institute of Nuclear Physics,
Moscow State University, RU-119991 Moscow, Russia
\vspace*{1ex}}
\affiliation{
Joint Institute for Nuclear Researches, RU-141980 Dubna, Russia
\vspace*{1ex}}
\author{S.V.~Petrushanko}
\affiliation{
Skobeltsyn Institute of Nuclear Physics,
Moscow State University, RU-119991 Moscow, Russia
\vspace*{1ex}}
\author{A.M.~Snigirev}
\affiliation{
Skobeltsyn Institute of Nuclear Physics,
Moscow State University, RU-119991 Moscow, Russia
\vspace*{1ex}}

\date{\today}

\begin{abstract}
Triangular flow $v_3$ of identified and inclusive particles in Pb+Pb 
collisions at $\sqrt{s_{NN}} = 2.76$~TeV is studied as a function of 
centrality and transverse momentum within the \textsc{hydjet++} model. 
The model enables one to investigate the influence 
of both hard processes and final-state interactions on the harmonics of 
particle anisotropic flow. Decays of resonances are found to increase 
the magnitude of the $v_3(p_{\rm T})$ distributions at $p_{\rm T} \geq 
2$~GeV/$c$ and shift their maxima to higher transverse momenta. The 
$p_{\rm T}$-integrated triangular flow, however, becomes slightly 
weakened for all centralities studied. The resonance decays also modify 
the spectra towards the number-of-constituent-quark scaling fulfillment 
for the triangular flow, whereas jets are the main source of the scaling 
violation at the energies available at the CERN Large Hadron Collider
(LHC). Comparison with the corresponding spectra of elliptic flow 
reveals that resonance decays and jets act in a similar manner on both 
$v_3(p_{\rm T})$ and $v_2(p_{\rm T})$ behavior. Obtained results are 
also confronted with the experimental data on differential triangular 
flow of identified hadrons, ratio $v_3^{1/3}(p_{\rm T}) / 
v_2^{1/2}(p_{\rm T})$ and $p_{\rm T}$-integrated triangular flow of
charged hadrons.
\end{abstract}

%
%
\maketitle
\section{Introduction}
\label{sec1}

Collective flow of hadrons produced in ultrarelativistic heavy-ion
collisions is one of the signals especially sensitive to the creation 
of even a small amount of quark-gluon plasma (QGP) 
\cite{SG_86,RR_97,BCLS_94}. To quantify this phenomenon an expansion 
of the azimuthal distribution of hadrons in a Fourier series was 
proposed in Refs.~\cite{VoZh96,PoVo98}:
\beq
\ds
\frac{d N}{d \phi} \propto 1 + 2 \sum \limits_{n=1}^{\infty} 
v_n \cos{\left[ n (\phi - \Psi_n) \right] }~.
\label{eq1}
\eeq
Here $\phi$ denotes the azimuthal angle between the particle transverse 
momentum and the participant event plane, and $\Psi_n$ is the azimuth 
of the event plane of the $n$th flow component. The coefficients $v_n$ 
are the flow harmonics that can be found after the averaging of cosines 
in Eq.~(\ref{eq1}) over all particles in an event and all events in the 
data sample: 
\beq
\ds
v_n = \llangle \cos{\left[ n (\phi - \Psi_n)\right] } \rrangle~.
\label{eq2}
\eeq
Modifications of the proposed analysis are possible \cite{CsSt14}, but 
we keep the traditional scheme to compare our results to 
the experimental data. For almost 20 years experimentalists and 
theorists have intensively investigated mainly the two lowest-order 
coefficients, called directed $v_1$ and elliptic $v_2$ flow, see, e.g. 
Refs.~\cite{VPS10,HeSn13} and references therein, whereas the study of 
triangular $v_3$ flow and higher harmonics started not long ago 
\cite{AlRo10,cms_11,atlas_11,alice_11}.

In collisions of similar nuclei, such as gold-gold or lead-lead, the 
higher-order odd-flow harmonics measured with respect to the reaction 
plane are expected to vanish under the assumption of a symmetric energy 
distribution. Experiments confirm that $v_3(\Psi_2) = 0$. However, 
initial-state fluctuations can lead to a non-vanishing participant 
triangularity \cite{AlRo10}, which is approximately linear to the 
triangular flow in its own participant plane $\Psi_3$ \cite{PQBM10,QH11}.
In many theoretical works devoted to the investigation of the signal in 
heavy-ion collisions, topics such as the response of $v_3$ to the initial 
triangularity $\varepsilon_3$ of the collision zone and sensitivity to 
initial-state fluctuations and to viscosity of hot QCD matter have been 
treated \cite{AlRo10,PQBM10,QH11,AGLO10}. Our study focuses mainly on 
the influence of jets and decays of resonances on the formation of $v_3$,
which to our best knowledge has not been explored extensively yet (see
Refs.~\cite{ST14} and \cite{QSH12}). The event generator 
\textsc{hydjet++} \cite{hydjet++} is employed for the simulation of
Pb+Pb collisions at $\sqrt{s_{NN}} = 2.76$~TeV.

\textsc{hydjet++} suits very well for these purposes because the model 
includes the treatment of both soft and hard processes and has an 
extensive table of hadronic resonances, including the charmed ones,
with more than 400 baryonic and mesonic states. The properties of the 
model and the generation of the anisotropic flow in it are discussed in 
Sec.~\ref{sec2}. Section~\ref{sec3} presents the results on 
differential and integrated triangular flow of charged hadrons produced
from central to (semi)peripheral Pb+Pb collisions at $\sqrt{s_{NN}} 
= 2.76$~TeV. Here the partial contributions of hydrodynamic processes, 
jets, and decays of resonances to the formation of the final $v_3$ are 
studied. Ratios $v_3^{1/3}/v_2^{1/2}$ and fulfillment of the 
number-of-constituent-quark (NCQ) scaling are investigated as well. 
Finally, conclusions are drawn in Sec.~\ref{sec4}.

\section{Generation of triangular flow in \textsc{hydjet++} }
\label{sec2}

The Monte Carlo event generator \textsc{hydjet++} (hydrodynamics with 
jets) treats a relativistic heavy-ion collision as a superposition of 
a soft, hydrolike state, hadronized as a result of a sudden thermal 
freeze-out, and a hard multiparton state, where energetic partons 
experience collisional and radiative energy losses in an expanding 
quark-gluon fluid \cite{hydjet++}. The simulation of both states 
proceeds independently. In the soft sector the thermalized system of 
hadrons is generated on the hypersurfaces of chemical and thermal 
freeze-out given by the parametrized relativistic hydrodynamics with 
preset freeze-out conditions \cite{fastmc1,fastmc2}. This approach is 
similar to the \textsc{therminator} model \cite{therm}. The effective 
thermal volume of the fireball is used to calculate the mean 
multiplicities of hadrons produced at the freeze-out hypersurface. The 
volume is generated on an event-by-event basis. It is proportional to 
the number of wounded nucleons at a given centrality provided by the 
Glauber model of multiparticle scattering. The only final-state 
interactions taken into account are the two- and three-body decays of 
resonances. The table of resonances is quite extensive and contains 
more than 360 meson and baryon (anti)states including the charmed ones.

The next part of \textsc{hydjet++}, which describes the hard partonic 
interactions, employs the generator \textsc{pyquen} \cite{pyquen} for 
simulation of single hard nucleon-nucleon collisions. It starts with 
the \textsc{pythia}-generated initial parton distributions and 
generation of the spatial vertexes of jet production, proceeds with the
rescattering-by-rescattering propagation of partons through the hot and
dense medium, determines the partons' mean free path as well as 
radiative and collisional energy loss, and, finally, hadronizes the hard
partons and in-medium emitted gluons according to the Lund string model.
Thus for each symmetric heavy-ion collision at a given impact parameter,
the mean number of jets is a product of binary $NN$ collisions and the 
integral cross section of the hard process with the certain minimum 
momentum transfer, $p_{\rm T}^{min}$. 
Partons produced in initial hard scatterings with a transverse momentum 
transfer lower than $p_{\rm T}^{min}$ are excluded from the hard 
component. Their products of hadronization are added to the thermalized 
component of the particle spectrum. These hadrons, however, can carry 
only weak anisotropic flow arising because of the well-known 
jet-quenching effect. Details of the model can be found in Refs. 
\cite{hydjet++,fastmc1,fastmc2,hj_epjc12}. It is worth noting that the 
combination of parametrized hydrodynamics with jets was able to explain 
the falloff of the elliptic flow, $v_2$, at high transverse momenta and 
violation of the mass ordering of $v_2(p_{\rm T})$ distributions for 
mesons and baryons at $p_{\rm T} \approx 2$~GeV/$c$ \cite{v2_prc09}, to 
predict the violation of the number-of-constituent-quark scaling for 
$v_2$ at the energies available at the LHC \cite{v2_prc09,jpg_sqm09}, 
and to describe the rise of the high-$p_{\rm T}$ tail of the $v_4/v_2^2$ 
ratio at the energies available at the BNL Relativistc Heavy Ion 
Collider (RHIC) and the LHC \cite{app_sqm11,v4_v2_prc13}. The extension 
of \textsc{hydjet++} to the triangular flow was done in 
Ref.~\cite{hj_epjc14}. The interplay of $v_2$ and $v_3$ in the model 
describes the non-linear contributions of elliptic and triangular flow 
to higher flow harmonics including the hexagonal one 
\cite{hj_epjc14,v6_prc14}, as well as the long-range dihadron 
correlations, known as ridge \cite{ridge_prc15}. Triangular flow is a 
very important ingredient of the model and it is thus worth discussing 
the features of its generation.

Anisotropic flow emerges in \textsc{hydjet++} because of the following
reasons. The profile of the nuclear overlap zone in the transverse 
plane can be approximated by an ellipse with the spatial eccentricity 
$\ds \varepsilon(b) = \frac{R_y^2 - R_x^2}{R_y^2 + R_x^2}$, where $b$ 
is the impact parameter, and $R_y$ and $R_x$ are the long- and 
short-ellipse radii, respectively. Then, the transverse radius of the 
fireball reads
\beq \ds
R_{\rm ell}(b, \phi) = R_{\rm fo}(b) \frac{\sqrt{1 - \varepsilon^2(b)}}
{\sqrt{1 + \varepsilon(b) \cos{2\phi}}}\ ,
\label{eq3}
\eeq
where 
\beq \ds
R_{\rm fo}(b) \equiv \sqrt{(R_x^2 + R_y^2)/2} = 
R_0 \sqrt{1 - \varepsilon(b)}
\label{eq4}
\eeq
with $R_0$ being the freeze-out transverse radius in a perfectly 
central collision with $b = 0$. Because every fluid cell is carrying a 
certain momentum, the spatial anisotropy at the freeze-out will be 
transformed into the momentum anisotropy. Unlike several other 
models, \textsc{hydjet++} does not rely on isotropic parametrization, 
where the azimuthal angle of the fluid velocity, $\phi_{\rm fluid}$, 
coincides with the azimuthal angle $\phi$. Instead, these two angles 
are correlated via the nonlinear function \cite{fastmc2}
\beq \ds
\tan{\phi_{\rm fluid}} = \sqrt{\frac{1 - \delta(b)}{1 + \delta(b)}}~
{\tan{\phi}}\ ,
\label{eq5}
\eeq
where the new anisotropy parameter, $\delta(b)$, is introduced. Both 
spatial and flow anisotropy parameters, $\varepsilon(b)$ and $\delta(b)$, 
are proportional to the initial spatial anisotropy $\varepsilon_0 = 
b/(2 R_A)$. Their values are fixed after fitting the \textsc{hydjet++} 
calculations to the measured elliptic flow. 

To extend the model to the triangular flow the transverse radius of the 
freeze-out hypersurface was modified accordingly \cite{hj_epjc14}:
\beq \ds
R_{\rm trian} (b,\phi) = R_{\rm ell}(b,\phi) \{ 1 + \varepsilon_3(b) 
\cos{\left[3(\phi - \Psi_3)\right]} \} \ .
\label{eq6}
\eeq
Because experiments show no correlation between the event planes of
the second and third harmonics, i.e., $v_2(\Psi_3) = v_3(\Psi_2) = 0$, 
these planes are also uncorrelated in \textsc{hydjet++}. The new
parameter $\varepsilon_3(b)$ entering Eq.~(\ref{eq6}) is responsible 
for the creation of triangularity in the system. Similarly to 
$\varepsilon(b)$, it can also be proportional to the initial 
eccentricity $\varepsilon_0(b)$ or considered as a free parameter.

Note, that the model possesses event-by-event (EbyE) fluctuations even
when the values of the anisotropy parameters, $\varepsilon(b), \delta(b)
\ {\rm and} \ \varepsilon_3(b)$, are fixed at the fixed impact parameter 
$b$. The sources of the EbyE fluctuations are fluctuations in particle 
multiplicities, coordinates, and momenta, as well as production of 
minijets and decays of resonances. Recently, \textsc{hydjet++} was 
extended \cite{ebye_epjc15} to match the measured EbyE fluctuations 
quantitatively. For this purpose the three parameters were not fixed 
anymore but rather smeared normally around their most probable values. 
The smearing procedure, however, does not change the distributions of 
either differential $v_{2(3)}(p_{\rm T},b)$ or $p_{\rm T}$-integrated 
$v_{2(3)}(b)$ characteristics, because these results are obtained after 
the averaging over quite substantial amounts of generated events.  

\section{Triangular flow in Pb+Pb collisions at LHC }
\label{sec3}

The transverse momentum dependence and the centrality dependence of 
the triangular flow of hadrons were studied in Pb+Pb collisions at 
center-of-mass energy $\sqrt{s_{NN}} = 2.76$~TeV. The considered 
$p_{\rm T}$ interval was $0 \leq p_{\rm T} \leq 8$~GeV/$c$, whereas 
the centrality range $0 \leq \sigma/\sigma_{\rm geo} \leq 50\%$ was
subdivided into five bins, namely, 0 - 10\%, 10 - 20\%, 20 - 30\%,
30 - 40\%, and 40 - 50\%. Because the yields of hadrons with transverse 
momenta larger than 1~GeV/$c$ rapidly drop, the generated event 
statistics was about 1 000 000 events for each centrality bin to 
provide reliable values for $v_3$ at $p_{\rm T} \geq 4$~GeV/$c$.
A detailed comparison of the \textsc{hydjet++} calculations of $v_3$ to 
the corresponding data by the ATLAS Collaboration \cite{atlas_prc12} and 
the CMS Collaboration \cite{cms_prc13} was done in Ref.~\cite{hj_epjc14}. 
In the present article our primary goal is to reveal the contributions of 
different processes to the formation of triangular flow. We start from
the interplay of soft processes and jets.
  
\subsection{Interplay of soft and hard processes }
\label{sec3_sub1}

\begin{figure}
\resizebox{\linewidth}{!}{
\includegraphics[scale=0.60]{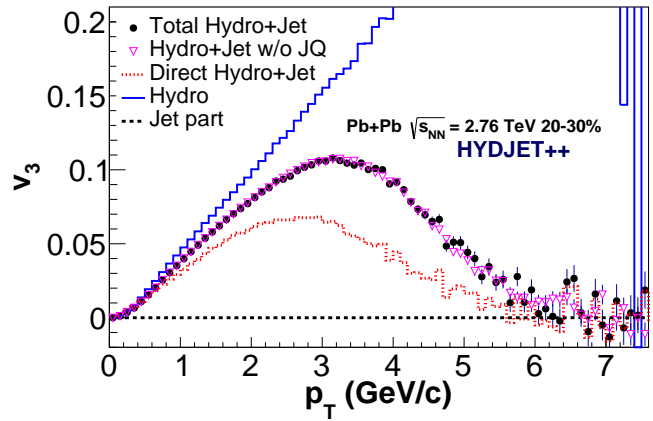}
}
\caption{(Color online)
The $p_{\rm T}$ dependence of different components of the triangular 
flow of charged particles produced in the \textsc{hydjet++} model for 
Pb + Pb collisions at $\sqrt{s_{NN}} = 2.76$~TeV at centrality 
20-30\%. The shown distributions are the $v_3(p_{\rm T})$ of particles 
coming from the hydro part (ii) (solid line), from the jets (iii) 
(dashed line), and of directly produced particles in both soft and hard 
(v) interactions (dotted line). Final triangular flow (vi) is presented 
by the solid circles, and triangles indicate the $v_3$ calculated 
without the jet quenching.
\label{fig1} }
\end{figure}

\begin{figure}
\resizebox{\linewidth}{!}{
\includegraphics[scale=0.60]{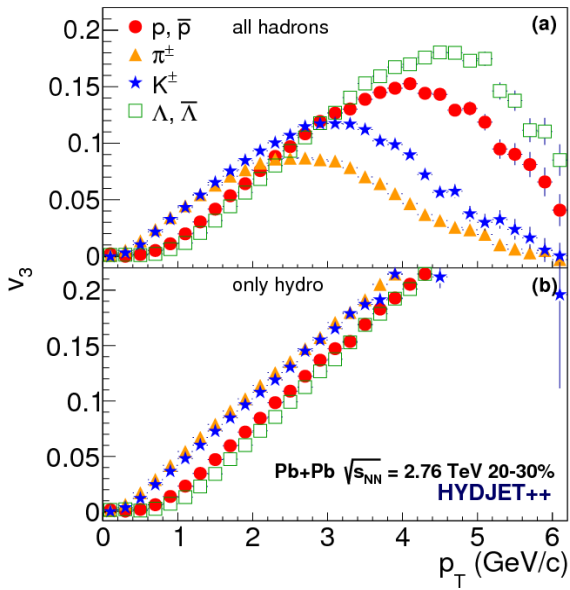}
}
\caption{(Color online)
The $p_{\rm T}$ dependence of (a) total triangular flow (vi) and (b) 
its hydro component (ii) in the \textsc{hydjet++} model for Pb + Pb 
collisions at $\sqrt{s_{NN}} = 2.76$~TeV at centrality 20-30\%. The 
hadron species are $p + \bar{p}$ (solid circles), charged pions (solid 
triangles), charged kaons (solid stars), and $\Lambda + 
\overline{\Lambda}$ (open squares).
\label{fig2} }
\end{figure}

In what follows, we will distinguish between the spectra of particles
\begin{itemize}
\item
(i) - directly frozen at the freeze-out hypersurface in hydro 
calculations (direct hydro),
\item
(ii) - direct hydro + resonance decays (soft processes only),
\item
(iii) - directly produced in jet fragmentation,
\item
(iv) - directly produced from jets + resonance decays (hard processes 
only),
\item
(v) - directly produced from hydro and jets, and
\item
(vi) - produced in all processes, i.e., hydro + jets + resonance decays.
\end{itemize}

Figure~\ref{fig1} shows the $v_3(p_{\rm T})$ spectrum of charged hadrons 
produced in collisions with centrality $20\% \leq \sigma/\sigma_{\rm geo} 
\leq 30\%$. In addition to the resulting triangular flow (vi), the 
partial contributions coming from the hydrodynamic part with resonances 
(ii), the jet fragmentation (iii), and particles produced either at the 
freeze-out hypersurface or decoupled from jets (v) are displayed as well. 
To study the influence of the jet-medium interaction on the triangular 
flow, we plot in Fig.~\ref{fig1} also the calculations without the jet 
quenching. Note that the triangular flow of hadrons originated from the 
jets both with and without the jet quenching is consistent with zero in 
the model. Hadrons coming from soft hydrodynamic processes demonstrate 
an almost linear rise of triangular flow with increasing transverse 
momentum at $0.3 \leq p_{\rm T} \leq 4$~GeV$/c$. Hydrodynamics, however, 
dominates the particle production at $p_{\rm T} \lesssim 2.5$~GeV/$c$ 
only, whereas at higher transverse momenta the particle spectrum is 
dominated by the jet hadrons. These circumstances cause the falloff of 
the $v_3(p_{\rm T})$ at $p_{\rm T} \geq 3.5$~GeV/$c$. 
The jet quenching enhances the yield of hadrons with low and 
intermediate transverse momenta. These particles should reduce the 
triangular flow in low- and intermediate-$p_{\rm T}$ ranges. However, 
their admixture is very small compared to hadrons produced in soft 
processes. As one can see in Fig.~\ref{fig1}, the impact of the jet 
quenching on the development of the $p_{\rm T}$-differential $v_3$ is 
insignificantly small. Decays of resonances increase 
the triangular flow of charged hadrons at $p_{\rm T} \geq 1$~GeV/$c$. 
This follows from the comparison of $v_3(p_{\rm T})$ of hadrons directly 
frozen at the freeze-out hypersurface or produced in the course of the 
jet fragmentation (dotted curve) with the total signal (solid circles), 
which includes also the hadrons coming from the decays of resonances.
The detailed discussion of the influence of resonance decays on the
triangular flow is given in Sec.~\ref{sec3_sub2}.

Transverse momentum distributions of the triangular flow (vi) of most 
abundant charged hadrons, such as pions, kaons, (anti)protons, and
(anti)$\Lambda$'s are depicted in Fig.~\ref{fig2}(a) together with the 
hydrodynamic parts (ii) of their spectra, shown separately in 
Fig.~\ref{fig2}(b). Several things are worth mentioning here. In 
hydrodynamic calculations both meson and baryon branches show a linear 
rise at $0.5 \leq p_{\rm T} \leq 5$~GeV$/c$. Mesonic flow is stronger 
than that of (anti)protons, whereas for full hydro+jets calculations 
this is true only for $p_{\rm T} \leq 2.5$~GeV/$c$. At higher transverse 
momenta the triangular flow of protons and antiprotons, $v_3^{\bar{p}+p}
(p_{\rm T})$, continues to rise, while the triangular flow of charged 
pions, $v_3^{\pi^\pm}(p_{\rm T})$, and kaons, $v_3^{K^\pm}(p_{\rm T})$, 
drops. This effect is also caused by the jet hadrons. The heavier the 
particle, the harder its $p_{\rm T}$-spectrum in hydrodynamics is. Thus, 
jets start to reduce the $v_3(p_{\rm T})$ distribution of heavy hadrons 
at larger transverse momenta compared to light hadrons, as seen in 
Fig.~\ref{fig2}(a).

\begin{figure}
\resizebox{\linewidth}{!}{
\includegraphics[scale=0.55]{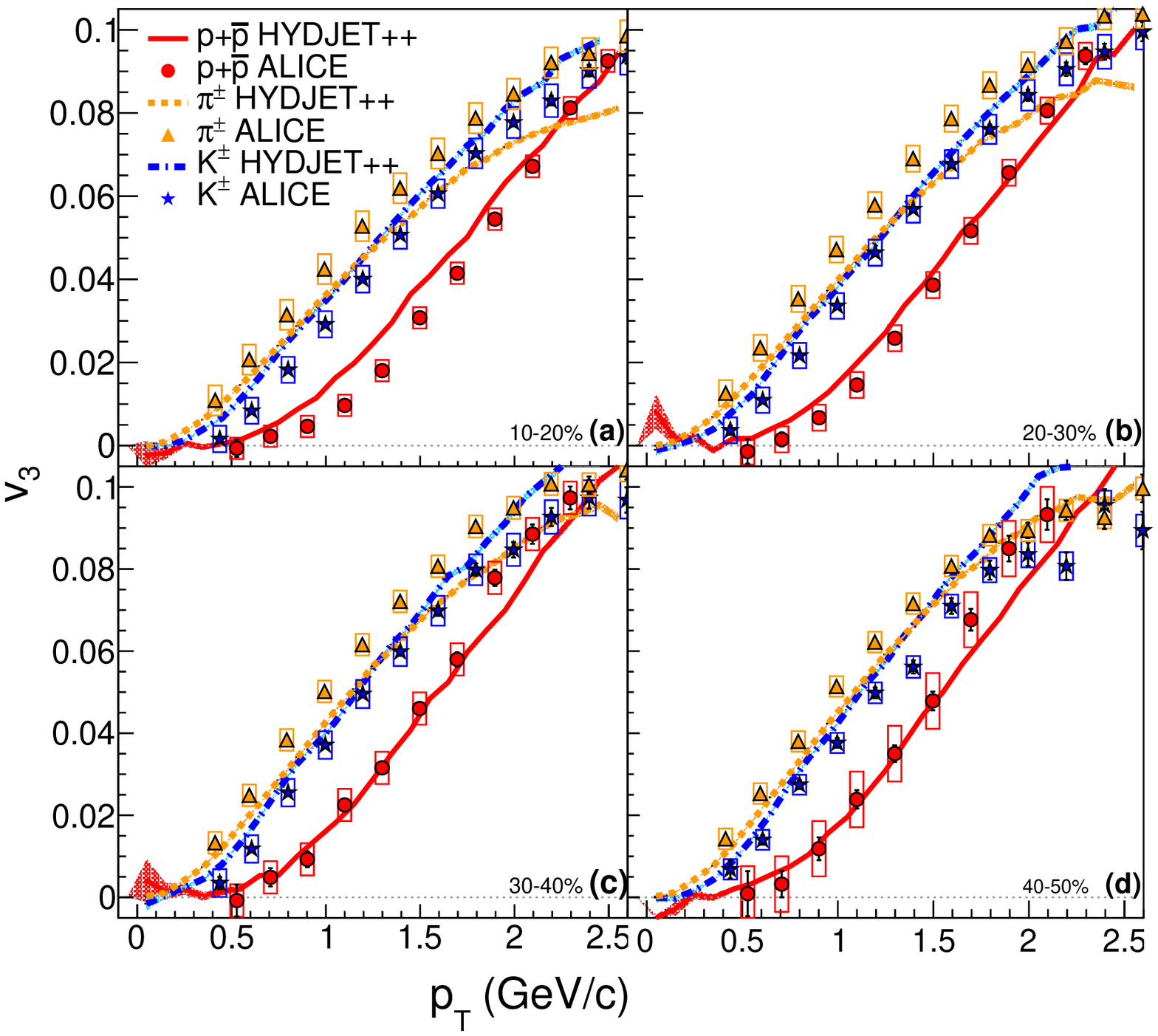}
}
\caption{(Color online)
The $p_{\rm T}$ dependence of triangular flow of $\pi^\pm$ (dashed 
lines), $K^\pm$ (dash-dotted lines) and $p + \bar{p}$ (solid lines) in
\textsc{hydjet++} calculations of Pb+Pb collisions at 2.76~TeV at 
centrality (a) 10-20\%, (b) 20-30\%, (c) 30-40\% and (d) 40-50\%.
Corresponding experimental data from Ref.~\cite{alice_v3_16} are shown 
by solid triangles ($\pi^\pm$), stars ($K^\pm$), and circles 
($p + \bar{p}$), respectively.
\label{fig3} }
\end{figure}

Recently, the triangular flow of identified hadrons in Pb+Pb collisions 
at $\sqrt{s} = 2.76$~TeV was measured at different centralities by the
ALICE Collaboration \cite{alice_v3_16}. The results of \textsc{hydjet++}
calculations for charged pions, kaons, and protons with antiprotons are
plotted onto the experimental data in Fig.~\ref{fig3} for four 
centrality bins, 10-20\%, 20-30\%, 30-40\%, and 40-50\%. It follows from
the comparison that the model provides a fair description of the data. 
It reproduces correctly the mass ordering of hadron $v_3$ at $p_{\rm T} 
\leq 2$~GeV/$c$ and its violation at higher transverse momenta.
     
\subsection{Influence of resonances}
\label{sec3_sub2}

\begin{figure}
\resizebox{\linewidth}{!}{
\includegraphics[scale=0.55]{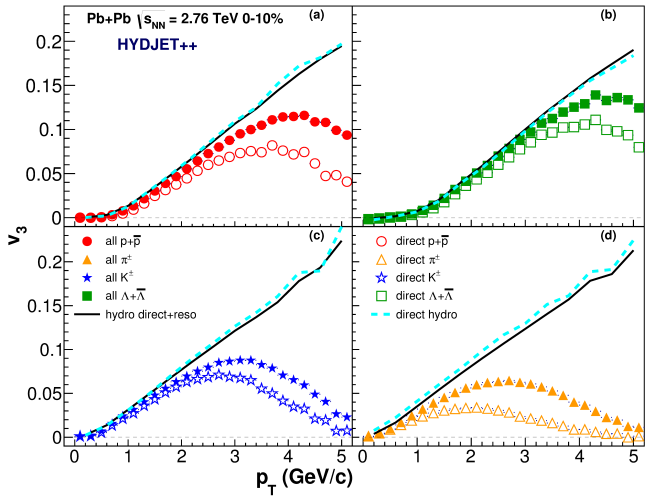}
}
\caption{(Color online)
The $p_{\rm T}$ dependence of triangular flow of (i) direct hadrons in
hydro (dashed lines), (ii) all hadrons in hydro (solid lines), (v) 
direct hadrons in soft and hard processes (open symbols), and (vi) all 
hadrons (solid symbols) produced in the \textsc{hydjet++} model for 
Pb + Pb collisions at $\sqrt{s_{NN}} = 2.76$~TeV with centrality 
$0 - 10\%$ for (a) $p + \bar{p}$, (b) $\Lambda + \overline{\Lambda}$, 
(c) charged kaons, and (d) charged pions.
\label{fig4} }
\end{figure}

\begin{figure}
\resizebox{\linewidth}{!}{
\includegraphics[scale=0.55]{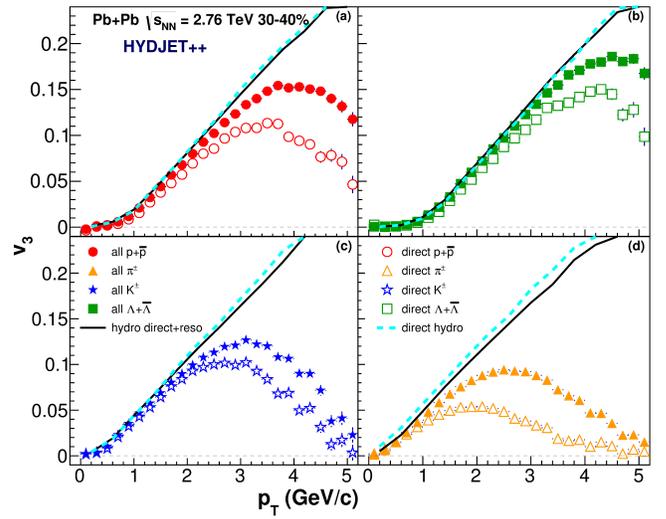}
}
\caption{(Color online)
The same as Fig.~\protect\ref{fig4} but for centrality $30 - 40\%$.
\label{fig5} }
\end{figure}

The differential spectra $v_3(p_{\rm T})$ of $\pi^\pm$, $K^\pm$, 
$p + \bar{p}$, and $\Lambda + \overline{\Lambda}$ are displayed in 
Fig.~\ref{fig4} for (semi)central (0 - 10\%) collisions and in 
Fig.~\ref{fig5} for semi-peripheral (30 - 40\%) collisions. Here the 
spectra of hadrons directly produced either on the freeze-out 
hypersurface or in the course of jet fragmentation (v) are compared to 
the final distributions (vi), where decays of resonances are taken into 
account. For both centrality intervals the physical picture is 
qualitatively similar. Namely, decays of resonances do not vary 
significantly the triangular flow of hadronic species at transverse 
momenta below 1~GeV/$c$ for pions and below 2~GeV/$c$ for heavier 
particles. At higher transverse momenta the situation is changed. 
Hadrons coming from the resonance decays enhance the differential 
$v_3$ of all species and shift the maxima of the distributions by 
0.5$-$1.0~GeV/$c$ towards higher $p_{\rm T}$. The maxima of 
$v_3(p_{\rm T})$ demonstrate the rise of about 25\% with 
shifting centrality from 0$-$10\% to 30$-$40\% (cf. Figs.~\ref{fig4} 
and \ref{fig5}).

In addition to these two distributions representing processes (v) and 
(vi), we plot in Figs.~\ref{fig4} and \ref{fig5} two curves showing
$v_3(p_{\rm T})$ of particles directly produced from hydro (i) and its
modification after the resonance decays (ii). For $p + \bar{p}$ and
$\Lambda + \overline{\Lambda}$ both curves are very close to each other,
whereas the $v_3(p_{\rm T})$ of charged kaons and, especially, of 
charged pions after the decays of resonances is a bit lower than that of 
directly produced particles. This result does not contradict to the
opposite behavior of the final spectra. The resonances are more 
abundantly produced in soft processes compared to the hard ones. Decays
of resonances significantly increase the particle yields in the soft 
part of the spectrum; therefore fractions of hydroparticles dominate
the particle spectra to larger transverse momenta. Consequently, the
rise of the $p_{\rm T}$-differential triangular flow will persist to 
larger values of $p_{\rm T}$. Recall that not all resonances are equally 
important. As was shown in Ref.~\cite{QSH12}, the set of resonances 
needed for the description of the flow harmonics $v_n$ of pions and 
protons can be reduced to 20$-$30 species only.

\begin{figure}
\resizebox{\linewidth}{!}{
\includegraphics[scale=0.55]{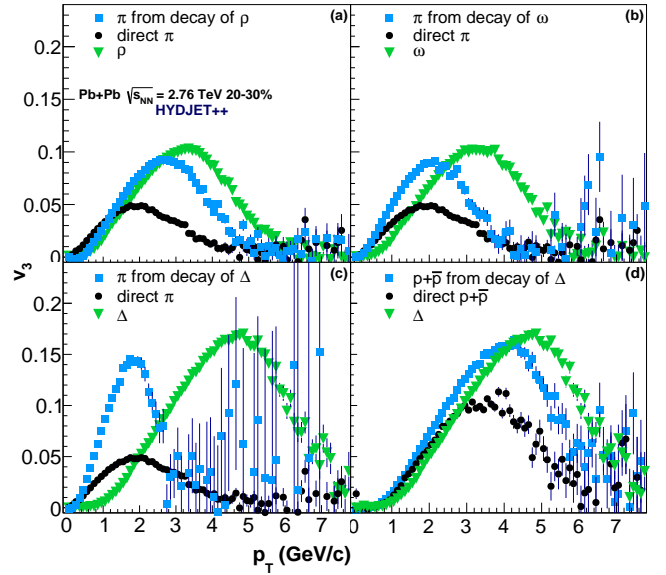}
}
\caption{(Color online)
The $p_{\rm T}$ dependence of triangular flow of charged pions produced 
both directly (circles) and in decays (squares) of (a) $\rho$ and (b) 
$\omega$, respectively, in the \textsc{hydjet++} model for Pb + Pb 
collisions at $\sqrt{s_{NN}}=2.76$~TeV with centrality $20 - 30\%$. 
(c) and (d) The same as panels (a) and (b) but for (c) charged pions 
and (d) $p + \bar{p}$ produced in decays of $\Delta$'s. The flow of 
resonances `is shown in each window by triangles.
\label{fig6} }
\end{figure}

On the other side, the main part of particle spectrum consists of 
hadrons with transverse momenta lower than 1~GeV/$c$ and pions are the 
dominant fraction of the spectrum, and for them the softest part of
the $v_3(p_{\rm T})$ distribution seems to carry a bit weaker triangular
flow after the decays of resonances compared to that of directly
produced pions, as shown in Fig.~\ref{fig5}. Therefore, the problem is
twofold. First, it is necessary to scrutinize how the decays of 
resonances alter the pion triangular flow. After that we should get the 
integrated values of $v_3$ at different centralities.

At the energies available at LHC of $\sqrt{s_{NN}}=2.76$~TeV or higher 
only about 20\% of pions are produced in \textsc{hydjet++} directly at 
the freeze-out hypersurface. The rest comes out as a result of decays of
various resonances, both mesonic and baryonic. If we consider an
isotropic decay of a baryon resonance on a pion and a lighter baryon,
then, because of the decay kinematics, the daughter baryon should
carry almost the same transverse momentum as the decaying resonance,
whereas the pion $p_{\rm T}$ is much softer. This type of reaction will
boost the triangular flow of the soft part of the pionic spectrum 
because the averaged triangular flow of heavy resonances is larger than 
that of pions. For meson resonances the softening or hardening of the 
pion triangular flow depends on the number of pions in the final state. 
To illustrate this let us consider three decays: $\rho \rightarrow \pi 
\pi$ (26\% of pion yield), $\omega \rightarrow \pi \pi \pi$ (11\%), and
$\Delta \rightarrow p + p/\bar{p}$ (less than 2\%). The differential
$v_3(p_{\rm T})$ of these resonances and their decay products are 
compared in Fig.~\ref{fig6} to the triangular flow of directly produced 
pions and (anti)protons. The triangular flow of pions from $\rho$ decays 
is just a bit softer than the $v_3$ of $\rho$ mesons. Consequently, it is 
harder than the triangular flow of direct pions at $p_{\rm T} \leq 
1.5$~GeV/$c$ [see Fig.~\ref{fig6}(a)]. The spectrum of pions from
$\omega$ decays, in contrast, is much softer than that of $\omega$'s.
Thus, their triangular flow at $p_{\rm T} \leq 1.5$~GeV/$c$ is even 
stronger than that of the direct pions. A similar tendency is revealed 
by pions from the $\Delta$ decays. Therefore, some resonances will 
enhance the pion triangular flow in the soft $p_{\rm T}$ region, whereas 
other will reduce it.

\begin{figure}
\includegraphics[width=0.48\textwidth]{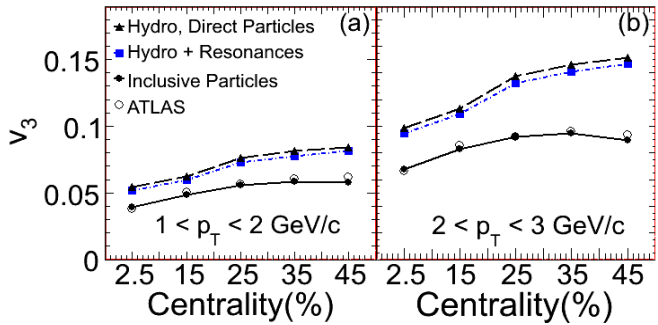}
\caption{(Color online)
$p_{\rm T}$-integrated triangular flow of inclusive charged hadrons
with (a) $1 \leq p_{\rm T} \leq 2$~GeV/$c$ and (b) $2 < p_{\rm T} \leq 
3$~GeV/$c$ as a function of centrality in Pb + Pb collisions at 
$\sqrt{s_{NN}} = 2.76$~TeV. Triangles and squares present the 
calculations for direct particles (i) and direct plus decays of 
resonances (ii), respectively, only in the hydro part of the model. 
Final results are shown by solid circles, and open circles are the 
experimental data from Ref.~\protect\cite{atlas_prc12}. Lines are drawn 
to guide the eye.
\label{fig7} }
\end{figure}

The result of this interplay is seen in Fig.~\ref{fig7}, which shows the 
integrated values of the triangular flow of charged hadrons calculated 
in five centrality bins. The triangular flow of direct particles and 
that of direct particles together with products of resonance decays 
obtained in hydro part of the model are shown separately. To compare the 
model results to the experimental data, the integration over the 
transverse momentum was done in two intervals:
$1 \leq p_{\rm T} \leq 2$~GeV/$c$, displayed in Fig.~\ref{fig7}(a), and 
$2 < p_{\rm T} \leq 3$~GeV/$c$, displayed in Fig.~\ref{fig7}(b). 
Experimental data from the ATLAS Collaboration \cite{atlas_prc12} are 
plotted onto the \textsc{hydjet++} calculations. We see that decays of 
resonances just slightly reduce the integrated $v_3$. The triangular 
flow at $2 < p_{\rm T} \leq 3$~GeV/$c$ in the hydrodynamic part is 
almost twice as stronger as the $v_3$ at $1 \leq p_{\rm T} \leq 
2$~GeV/$c$. Jets significantly diminish the triangular flow in both 
$p_{\rm T}$ intervals. 

\subsection{Ratio $v_3^{1/3} / v_2^{1/2}$}
\label{sec3_sub3}

\begin{figure}
\includegraphics[width=0.48\textwidth]{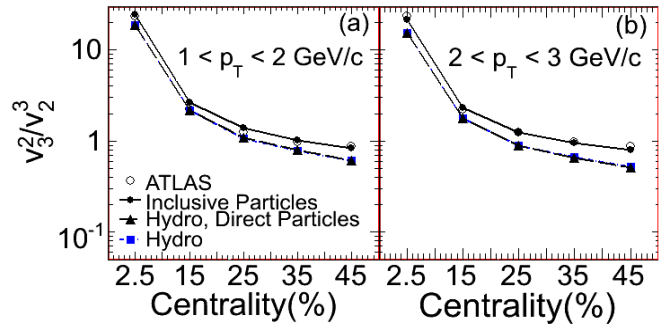}
\caption{(Color online)
The same as Fig.~\protect\ref{fig7} but for the ratio $v_3^2 / v_2^3$ 
of $p_{\rm T}$-integrated triangular and elliptic flow.
\label{fig8} }
\end{figure}

The ratios $v_n^{1/n} / v_2^{1/2}$ were suggested in 
Ref.~\cite{atlas_prc12} as a probe to study the possible scaling 
properties of the flow harmonics. The ratio of triangular and elliptic 
flow, where both harmonics are integrated over the transverse momentum 
range, reveals no indication of the scaling trend. The results are 
presented in Fig.~\ref{fig8}. As in the previous figure, two groups of 
hadrons are selected, one with $1 \leq p_{\rm T} \leq 2$~GeV/$c$ and 
another with $2 < p_{\rm }T \leq 3$~GeV/$c$ to compare the 
\textsc{hydjet++} calculations to the experimental data. To see the 
changing of the ratio with increasing impact parameter more distinctly, 
the ratio $v_3^2/v_2^3$ is used. Again, the decays of resonances make no 
impact on the ratio, which clearly drops as the reaction becomes more 
peripheral. Note that jets increase the final ratio compared to the pure 
hydro part. We should come back to this point later.
 
\begin{figure}
\resizebox{\linewidth}{!}{
\includegraphics[scale=0.55]{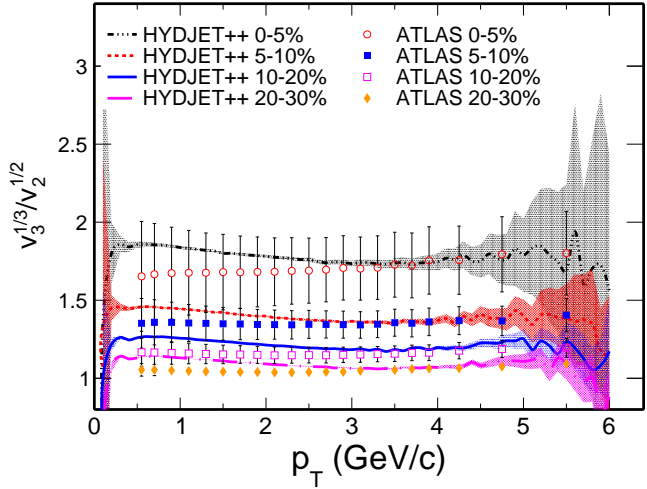}
}
\caption{(Color online)
Ratio $v_3^{1/3} / v_2^{1/2}$ as a function of $p_{\rm T}$ in centrality 
bins 0-5\% (dotted line), 5-10\% (dashed line), 10-20\% (solid line), 
and 20-30 \% (dash-dotted line). Shaded areas indicate the statistical 
error bands in the model. The corresponding ATLAS data from Ref. 
\protect\cite{atlas_prc12} are shown by open circles, filled squares, 
open squares, and diamonds, respectively. 
\label{fig9} }
\end{figure}

If the ratio $v_3^{1/3}/v_2^{1/2}$ is plotted as a function of 
transverse momentum in various centrality bins, as shown in 
Fig.~\ref{fig9}, then the considered distributions are remarkably flat.
The ratios $v_3^{1/3}(p_{\rm T})/v_2^{1/2}(p_{\rm T})$ do not depend on 
$p_{\rm T}$ in a quite broad range of transverse momentum from 1~GeV/$c$ 
up to 6~GeV/$c$. This $p_{\rm T}$ independence, however, is not 
predefined in \textsc{hydjet++}, but arises as a result of nontrivial 
interplay between soft and hard processes. 

\begin{figure}
\resizebox{\linewidth}{!}{
\includegraphics[scale=0.70]{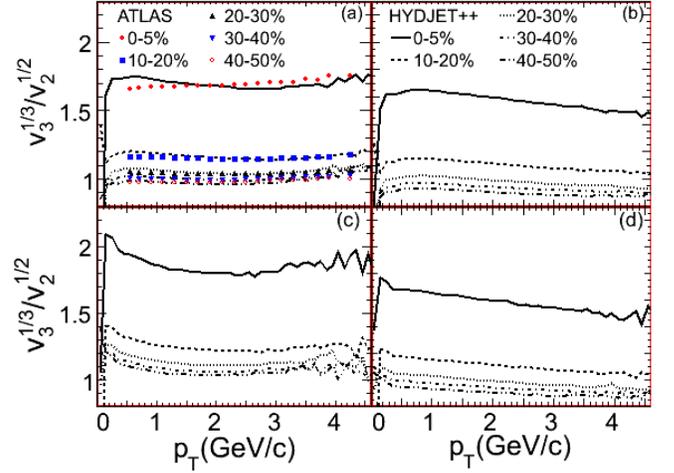}
}
\caption{(Color online)
The same as Fig.~\protect\ref{fig9} but for (a) ratios of total signals 
in $0-5\%$ (solid curve), $10-20\%$ (dashed curve), $20-30\%$ (dotted 
curve), $30-40\%$ (dash-dotted curve), and $40-50\%$ (dash-dot-dotted 
curve) centrality bins;
(b) ratios of only hydrodynamic parts (ii) of both flows; (c) ratios of 
the flow harmonics for only directly produced particles (v); and (d) 
ratios of the flow harmonics for only directly produced particles in 
the hydrodynamic part of the model (i).   
\label{fig10} }
\end{figure}

To study this effect we selected from the particle spectrum the 
following types of hadrons: directly produced hadrons (v), hadrons 
directly produced in the soft processes only (i), and hadrons from 
direct hydro plus the hadrons from resonance decays (ii). The calculated 
ratios $v_3^{1/3}(p_{\rm T}) / v_2^{1/2}(p_{\rm T})$ are depicted in 
Fig.~\ref{fig10}. In Fig.~\ref{fig10}(d) one can see that this ratio 
decreases with rising $p_{\rm T}$ for directly produced hadrons in the 
hydromodulus of the model for all centrality intervals. Decays of 
resonances make the slopes of the ratios less steep [see 
Fig.~\ref{fig10}(b)]. Finally, the rise of the tails at $p_{\rm T} \geq 
2$~GeV/$c$ is provided by the jet particles, as shown in 
Fig.~\ref{fig10}(c). Here the jets and the final-state interactions are 
working together towards the formation of a plateau at $ 1 \leq 
p_{\rm T} \leq 6$~GeV/$c$. At higher transverse momenta jets may cause 
the rise of the ratio, similar to that of $v^{1/4}/v_2^{1/2}$, observed 
both experimentally \cite{alice_v4v2} and in \textsc{hydjet++} 
\cite{v4_v2_prc13}. The lack of statistics and large error bars, 
however, does not permit us to make more definite conclusions.      

\subsection{NCQ scaling}
\label{sec3_sub4}

The NCQ scaling was first observed for the elliptic flow of hadron 
species in Au+Au collisions at the energy available at the RHIC of
$\sqrt{s_{NN}} = 200$~GeV \cite{star_ncq,phenix_ncq}. It was found that 
if one plots the $v_2$ as a function of the transverse kinetic energy 
of the hadron, $K E_{\rm T} \equiv m_{\rm T} - m_0$, and divides both 
$v_2$ and $K E_{\rm T}$ by the number of constituent quarks in the 
hadron, $n_q$, then the excitation functions $v_2^{hadr}(K E_{\rm T}/
n_q) / n_q$ of different hadrons sit on the top of each other up to 
$K E_{\rm T} / n_q \approx 0.8 \div 1.0$~GeV \cite{phenix_ncq_07}. It 
was pointed out in Refs.~\cite{v2_prc09,jpg_sqm09} that, because of the 
stronger jet influence, the fulfillment of the NCQ scaling at the LHC 
should be worsened compared to that at the RHIC. The worsening of the 
NCQ scaling conditions for the elliptic flow at the LHC was later 
observed by the ALICE Collaboration \cite{alice_ncq_1,alice_ncq_2}. 

\begin{figure}[t]
\resizebox{\linewidth}{!}{
\includegraphics[scale=0.65]{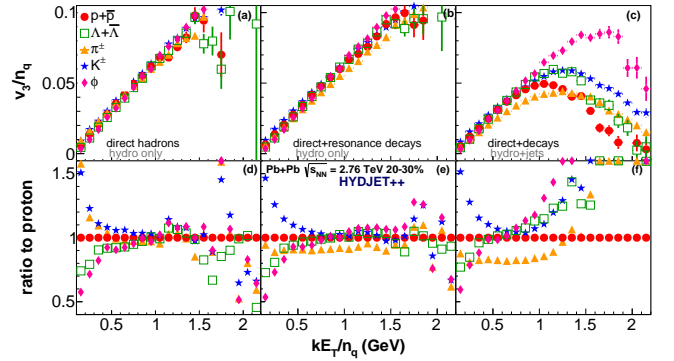}
}
\caption{(Color online)
Upper row: The $K E_{\rm T} / n_q$ dependence of the triangular flow for 
(a) direct hadrons (i), (b) hadrons produced in soft processes only (ii), 
and (c) hadrons produced both in soft and hard processes (vi) in the 
\textsc{hydjet++} model in Pb+Pb collisions at $\sqrt{s_{NN}} = 
2.76$~TeV with centrality $20-30\%$. The considered hadron species are:
$p + \bar{p}$ (solid circles), $\Lambda + \overline{\Lambda}$ (open
squares), $\pi^\pm$ (solid triangles), $K^\pm$ (stars), and $\phi$
(diamonds). Bottom row: The $K E_T / n_q$ dependence of the 
distributions in the upper row normalized to the triangular flow of 
$p + \bar{p}$, $(v_3 / n_q) / (v_3^{p + \bar{p}} /3)$. 
\label{fig11} }
\end{figure}

It is instructive, therefore, to check the NCQ scaling for the 
triangular flow of hadronic species at $\sqrt{s_{NN}} = 2.76$~TeV.
To elaborate on the role of final-state interactions and the hard
processes, we plot in Fig.~\ref{fig11} separately (a) the triangular 
flow of the main hadron species produced directly on the freeze-out
hypersurface (i), (b) then added to their spectra the flow of particles
produced after the decays of resonances (ii), and finally (c) the 
resulting $v_3$ of hadrons produced in both soft and hard processes 
(vi). For clarity, all distribution functions $v_3^{hadr} (K E_{\rm T}/ 
n_q) / n_q$ were also normalized to that of (anti)protons, shown in the 
bottom row of Fig.~\ref{fig11}. One can see in Fig.~\ref{fig11}(a) that 
the NCQ scaling is fulfilled in \textsc{hydjet++} within the 10\% 
accuracy limit for the $v_3$ of main hadron species, frozen already at 
the freeze-out hypersurface, in the range $0.5 \leq K E_{\rm T} \leq 
1.2$~GeV. This occurs because, as we already saw in Figs.~\ref{fig4} and 
\ref{fig5}, resonances increase the triangular flow of lighter hadrons 
at intermediate $p_{\rm T}$ and shift the maxima of their differential 
distributions to higher $p_{\rm T}$ values. Some hadrons, such as
$\phi$ mesons, do not get the feed-down from resonances, thus their 
distributions become closer to those of light mesons at intermediate 
transverse momenta. However, at $p_{\rm T} \gtrsim 3$~GeV/$c$ the 
particle spectra are dominated by the jet hadrons, for which the scaling 
conditions are not relevant. The hadrons fragmenting from jets lead to 
only approximate fulfillment of the NCQ scaling for the hadron 
triangular flow in the interval $0.15 \leq K E_{\rm T} \leq 1.1$~GeV.

\begin{figure}
\resizebox{\linewidth}{!}{
\includegraphics[scale=0.65]{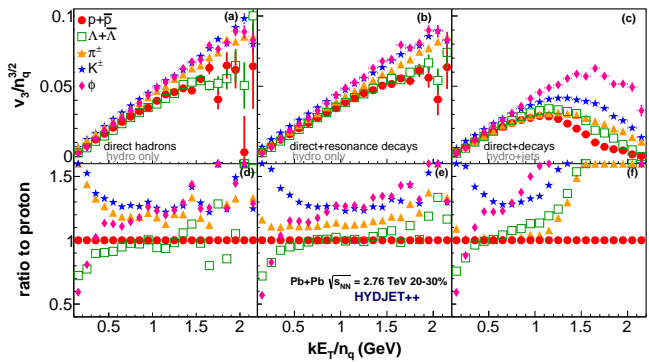}
}
\caption{(Color online)
The same as Fig.~\protect\ref{fig11} but for 
$v_3(K E_T /n_q) / n_q^{3/2}$ distributions.
\label{fig12} }
\end{figure}

It was suggested in Ref.~\cite{lacey_jpg11} to use the ratios 
$v_n^{hadr} / (n_q)^{n/2}$ instead of the standard $v_n^{hadr} / n_q$ 
to search for the NCQ scaling of the $n$th flow harmonic. The modified 
scaling for $0-50\%$ central Au+Au collisions at the highest energy 
available at the RHIC was observed for $v_2, v_3$ and $v_4$ 
\cite{phenix_mod_ncq}.
The \textsc{hydjet++} distributions $v_3^{hadr} / n_q^{3/2}$ are shown 
in Fig.~\ref{fig12}. Only approximate scaling within $\pm 15\%$ margins 
is seen for hadrons with $0.5 \leq KE_T \leq 1.5$~GeV produced in soft 
processes. When jets are taken into account, the interval of approximate 
scaling fulfillment for all but $\phi$ mesons shrinks to 
$0.5 \leq KE_T \leq 1.0$~GeV.  
          
\section{Conclusions}
\label{sec4}

The triangular flow $v_3$ of charged inclusive and identified hadrons 
was studied within the \textsc{hydjet++} model in Pb+Pb collisions at 
$\sqrt{s_{NN}} = 2.76$~TeV and centralities 
$0\% \leq \sigma/\sigma_{\rm geo} \leq 50\%$. The model couples soft
hydro-like state to hard processes and contains an extended table of
resonances, thus allowing for investigation of the interplay between 
soft processes, jets, and resonance decays on the formation of 
particle $v_3$. The results can be summarized as follows.

The triangular flows of identified hadrons produced in soft processes 
display an almost linear rise at $0.3 \leq p_{\rm T} \leq 5$~GeV/$c$. 
The mass ordering effect is achieved, i.e., the flow of mesons is 
stronger than that of baryons. The fraction of hadrons produced in jet 
fragmentation is the most abundant in particle spectra at $p_{\rm T} 
\geq 2.5$~GeV/$c$.  Since these hadrons carry almost no $v_3$, the 
distribution functions $v_3(p_{\rm T})$ experience a falloff at 
intermediate transverse momenta. The interplay of hard and soft 
processes leads also to breaking of the mass ordering of the triangular 
flow, because jet particles start to dominate spectra of heavy hadrons 
at larger $p_{\rm T}$ compared to those of light hadrons.
It appears that switching off the jet quenching does not influence the
final differential triangular flow $v_3(p_{\rm T})$. Model calculations
agree well with recent experimental data.

Decays of resonances distinctly modify the differential distributions
of hadrons $v_3(p_{\rm T})$ at $p_{\rm T} \geq 2$~GeV/$c$. The maxima of 
the spectra become about 25\% higher. Simultaneously, they are shifted 
by $0.5 - 1.0$~GeV/$c$ towards higher transverse momenta. In contrast,
the influence of resonance decays on the $p_{\rm T}$-integrated 
triangular flow is extremely small.  

The flatness of the ratios $v_3^{1/3} (p_{\rm T}) / v_2^{1/2} 
(p_{\rm T})$ at different centralities emerges in \textsc{hydjet++} as a 
result of interplay of final-state interactions and jets. These ratios 
decrease with rising transverse momenta for particles directly frozen at 
the freeze-out hypersurface. Decays of resonances reduce the values of
$v_3^{1/3} / v_2^{1/2}$ at low $p_{\rm T}$, whereas the jet hadrons 
boost their high-$p_{\rm T}$ tails, thus leading to independence of the 
ratios on transverse momentum in a broad range of $0.5 \leq p_{\rm T} 
\leq 5$~GeV/$c$.

The two mechanisms, however, work in opposite directions when we
consider the fulfillment of the NCQ scaling for the triangular flow. In 
this case decays of resonances enhance the high-$p_{\rm T}$ parts of the 
$v_3^{hadr}(K E_{\rm T}/n_q)/n_q$ spectra of light hadrons, thus 
extending the upper $K E_{\rm T}$ limit of the NCQ scaling performance. 
Jet particles, in their turn, carry very weak flow and wash out the 
signal. We verified also the NCQ scaling conditions for $v_3^{hadr}
(K E_{\rm T}/n_q)/n_q^{3/2}$ distributions. The result stays put; i.e., 
hadrons decoupling from jets are worsening the scaling, while the 
final-state interactions act toward its fulfillment. 

\begin{acknowledgments}
Fruitful discussions with M.~Bleicher, L.~Csernai, I.~Mishustin, and 
H.~St\"{o}cker are gratefully acknowledged. We thank also our colleagues 
from the CMS, ALICE, and ATLAS Collaborations for the extended cooperation.
This work was supported in parts by Grant No. INGO II LG15001 of the 
Ministry of Education, Youth and Sports of the Czech Republic; 
the Grant Agency of Czech Republic under Grant No. 13-20841S; 
the Department of Physics, University of Oslo;
the Frankfurt Institute for Advanced Studies (FIAS); and
the grant from the President of Russian Federation
for Scientific Schools (Grant No. 7989.2016.2).
L.B. acknowledges the financial support of the Alexander von Humboldt 
Foundation. B.H.B.J. acknowledges the financial support of
the Norwegian Research Council (NFR).
\end{acknowledgments}

\end{document}